\documentclass[preprint, 4p]{elsarticle}
\usepackage[utf8]{inputenc}

\usepackage{amssymb}

\usepackage{amsmath}
\usepackage{lineno}

\usepackage{subcaption}
\usepackage{hyperref}

\journal{SoftwareX}

\RequirePackage{listings}
\RequirePackage{xcolor}

\definecolor{gray}{gray}{0.5}
\colorlet{commentcolour}{green!50!black}

\colorlet{stringcolour}{red!60!black}
\colorlet{keywordcolour}{magenta!90!black}
\colorlet{exceptioncolour}{yellow!50!red}
\colorlet{commandcolour}{blue!60!black}
\colorlet{numpycolour}{blue!60!green}
\colorlet{literatecolour}{magenta!90!black}
\colorlet{promptcolour}{green!50!black}
\colorlet{specmethodcolour}{violet}

\newcommand*{\literatecolour}{\textcolor{literatecolour}}

\newcommand*{\pythonprompt}{\textcolor{promptcolour}{{>}{>}{>}}}

\lstdefinestyle{mypython}{
	language=python,
	showtabs=true,
	tab=,
	tabsize=2,
	basicstyle=\ttfamily\footnotesize,%\setstretch{.5},
	stringstyle=\color{stringcolour},
	showstringspaces=false,
	alsoletter={1234567890},
	otherkeywords={\%, \}, \{, \&, \|},
	keywordstyle=\color{keywordcolour}\bfseries,
	emph={and,break,class,continue,def,yield,del,elif ,else,%
		except,exec,finally,for,from,global,if,import,in,%
		lambda,not,or,pass,print,raise,return,try,while,assert,with},
	emphstyle=\color{blue}\bfseries,
	emph={[2]True, False, None},
	emphstyle=[2]\color{keywordcolour},
	emph={[3]object,type,isinstance,copy,deepcopy,zip,enumerate,reversed,list,set,len,dict,tuple,xrange,append,execfile,real,imag,reduce,str,repr},
	emphstyle=[3]\color{commandcolour},
	emph={Exception,NameError,IndexError,SyntaxError,TypeError,ValueError,OverflowError,ZeroDivisionError},
	emphstyle=\color{exceptioncolour}\bfseries,
	morecomment=[s]{"""}{"""},
	commentstyle=\color{commentcolour}\slshape,
	emph={[4]ode, fsolve, sqrt, exp, sin, cos,arctan, arctan2, arccos, pi,  
		array, norm, solve, dot, arange, isscalar, max, sum, flatten, shape, 
		reshape, find, any, all, abs, plot, linspace, legend, quad, 
		polyval,polyfit, hstack, 
		concatenate,vstack,column_stack,empty,zeros,ones,rand,vander,grid,pcolor,eig,eigs,eigvals,svd,qr,tan,det,logspace,roll,min,mean,cumsum,cumprod,diff,vectorize,lstsq,cla,eye,xlabel,ylabel,squeeze},
	emphstyle=[4]\color{numpycolour},
	emph={[5]__init__,__add__,__mul__,__div__,__sub__,__call__,__getitem__,__setitem__,__eq__,__ne__,__nonzero__,__rmul__,__radd__,__repr__,__str__,__get__,__truediv__,__pow__,__name__,__future__,__all__},
	emphstyle=[5]\color{specmethodcolour},
	emph={[6]assert,yield},
	emphstyle=[6]\color{keywordcolour}\bfseries,
	emph={[7]range},
	emphstyle={[7]\color{keywordcolour}\bfseries},
	literate=*%
	{:}{{\literatecolour:}}{1}%
	{=}{{\literatecolour=}}{1}%
	{-}{{\literatecolour-}}{1}%
	{+}{{\literatecolour+}}{1}%
	{*}{{\literatecolour*}}{1}%
	{**}{{\literatecolour{**}}}2%
	{/}{{\literatecolour/}}{1}%
	{//}{{\literatecolour{//}}}2%
	{!}{{\literatecolour!}}{1}%
	{[}{{\literatecolour[}}{1}%
	{]}{{\literatecolour]}}{1}%
	{<}{{\literatecolour<}}{1}%
	{>}{{\literatecolour>}}{1}%
	{>>>}{\pythonprompt}{3}%
	,%
	frame=trbl,
	rulecolor=\color{black!40},
	backgroundcolor=\color{white},
	breakindent=.5\textwidth,frame=single,breaklines=true%
}

\lstnewenvironment{python}[1][]{\lstset{style=mypython}}{}

\lstdefinestyle{mypythoninline}{
	style=mypython,%
	basicstyle=\ttfamily,%
	keywordstyle=\color{keywordcolour},%
	emphstyle={[7]\color{keywordcolour}},%
	emphstyle=\color{exceptioncolour},%
	literate=*%
	{:}{{\literatecolour:}}{2}%
	{=}{{\literatecolour=}}{2}%
	{-}{{\literatecolour-}}{2}%
	{+}{{\literatecolour+}}{2}%
	{*}{{\literatecolour*}}2%
	{**}{{\literatecolour{**}}}3%
	{/}{{\literatecolour/}}{2}%
	{//}{{\literatecolour{//}}}{2}%
	{!}{{\literatecolour!}}{2}%
	{[}{{\literatecolour[}}{2}%
	{]}{{\literatecolour]}}{2}%
	{<}{{\literatecolour<}}{2}%
	{<=}{{\literatecolour{<=}}}3%
	{>}{{\literatecolour>}}{2}%
	{>=}{{\literatecolour{>=}}}3%
	{==}{{\literatecolour{==}}}3%
	{!=}{{\literatecolour{!=}}}3%
	{+=}{{\literatecolour{+=}}}3%
	{-=}{{\literatecolour{-=}}}3%
	{*=}{{\literatecolour{*=}}}3%
	{/=}{{\literatecolour{/=}}}3%
}

\newcommand*{\pyth}{\lstinline[style=mypythoninline]}

\begin{document}

\begin{frontmatter}

\title{zfit: scalable pythonic fitting}

\author{Jonas Eschle}
\ead{Jonas.Eschle@cern.ch}
\author{Albert Puig Navarro}
\ead{albert.puig.navarro@gmail.com}
\author{Rafael Silva Coutinho}
\ead{rafael.silva.coutinho@cern.ch}
\author{Nicola Serra}
\ead{nicola.serra@cern.ch}

\address{Physik-Institut, Universität Zürich, Zürich (Switzerland)}

\begin{abstract}
Statistical modeling is a key element in many scientific fields and especially in High-Energy Physics (HEP) analysis.
The standard framework to perform this task in HEP is
the C++ ROOT/RooFit toolkit; with Python
bindings that are only loosely integrated into the scientific Python ecosystem.
In this paper, zfit, a new alternative to RooFit written in pure Python, is presented.
Most of all, zfit provides a well defined high-level API and workflow for advanced model building and fitting, together with an implementation on top of TensorFlow, allowing a transparent usage of CPUs and GPUs.
It is designed to be extendable in a very simple fashion, allowing the usage of cutting-edge developments from the scientific Python ecosystem in a transparent way.
The main features of zfit are introduced, and its extension to data analysis, especially in the context of HEP experiments, is discussed.

\end{abstract}


\end{frontmatter}

% \linenumbers

%\input{motivation}
%\input{description}
%\input{examples}
%\input{impact}

%\input{acknowledgements}

%\input{appendix}

\newcommand{\fig}[1]{Fig.~\ref{fig:#1}}

\section{Introduction\label{sec:motivation}}

Data collected by experiments such as the Large Hadron Collider (LHC) at CERN or ultra-energetic cosmic ray detectors are analysed in terms of physics-motivated mathematical models.
Typically, these models have a parametric form, and the values of their parameters are the quantities of interest in a given analysis.
Once a model has been defined, a \emph{fitting} procedure that minimizes the disagreement between the data and the model is applied to find the optimal values of these parameters, a crucial step in almost every analysis.

Model fitting as done in High Energy Physics (HEP) has some peculiarities and differs from other fields. Models are often multidimensional, built by composing complicated, custom shapes, and have a non-trivial normalization.
In addition, two further requirements need to be fulfilled:
on one hand, reasonable scaling with data size and model complexity, motivated by the large data samples being collected at the LHC; and, on the other hand, the capability to not only find optimal values for the parameters, but to be able to calculate their uncertainties with enough flexibility taking all correlations into account to allow for advanced statistical treatment, as uncertainties of measurements are as important as their central values.

Since recent years, HEP analysis is moving more and more towards the usage of Python over the more traditional C++.
The moving force behind this paradigm shift is the existence of an extremely rich scientific computing Python ecosystem~\cite{Oliphant:2007,Millman:2011}, which provides a collection of efficient, easy-to-use tools for data analysis that are shared across a very large community, including scientists from various fields and industry, especially in data science, and which one can build new applications on.

The \emph{de facto} standard for model fitting in HEP is RooFit~\cite{RooFit}, which is integrated in the ROOT toolkit~\cite{ROOT} and provides a very powerful object-oriented architecture, as well as plotting capabilities and an advanced statistics module.
It is written in C++ and can be accessed in Python through the ROOT Python bindings. These bindings come with its own set of non-negligible problems:
complex memory management that easily leads to memory leaks, difficult integration with the scientific computing Python stack, lack of extendability from Python, and a complex installation procedure due to the need to install the full ROOT toolkit.

General-purpose Python-based libraries such as SciPy~\cite{scipy}, lmfit~\cite{lmfit} or TensorFlow Probability~\cite{tfp} do not fulfill all HEP fitting requirements, and therefore cannot be used \emph{as-is}.
Python-based HEP-specific fitting packages, such as probfit~\cite{probfit}, TensorProb~\cite{tensorprob} or TensorFlowAnalysis~\cite{tfa}, have tried to address these issues and can be considered as proof-of-concepts that HEP-centric model fitting is possible in Python;
none of them, however, offers the same fitting functionalities and ease of use as RooFit does while still integrating well with the scientific computing Python ecosystem.

The goal of zfit is to fill this important gap in the HEP software ecosystem and provide a well defined API and workflow for model fitting together with a Python-based package that is fast, scalable and extensible.

\section{Software description\label{sec:sw-description}}

\begin{figure}[htb]
    \centering
    \includegraphics[width=1.0\linewidth]{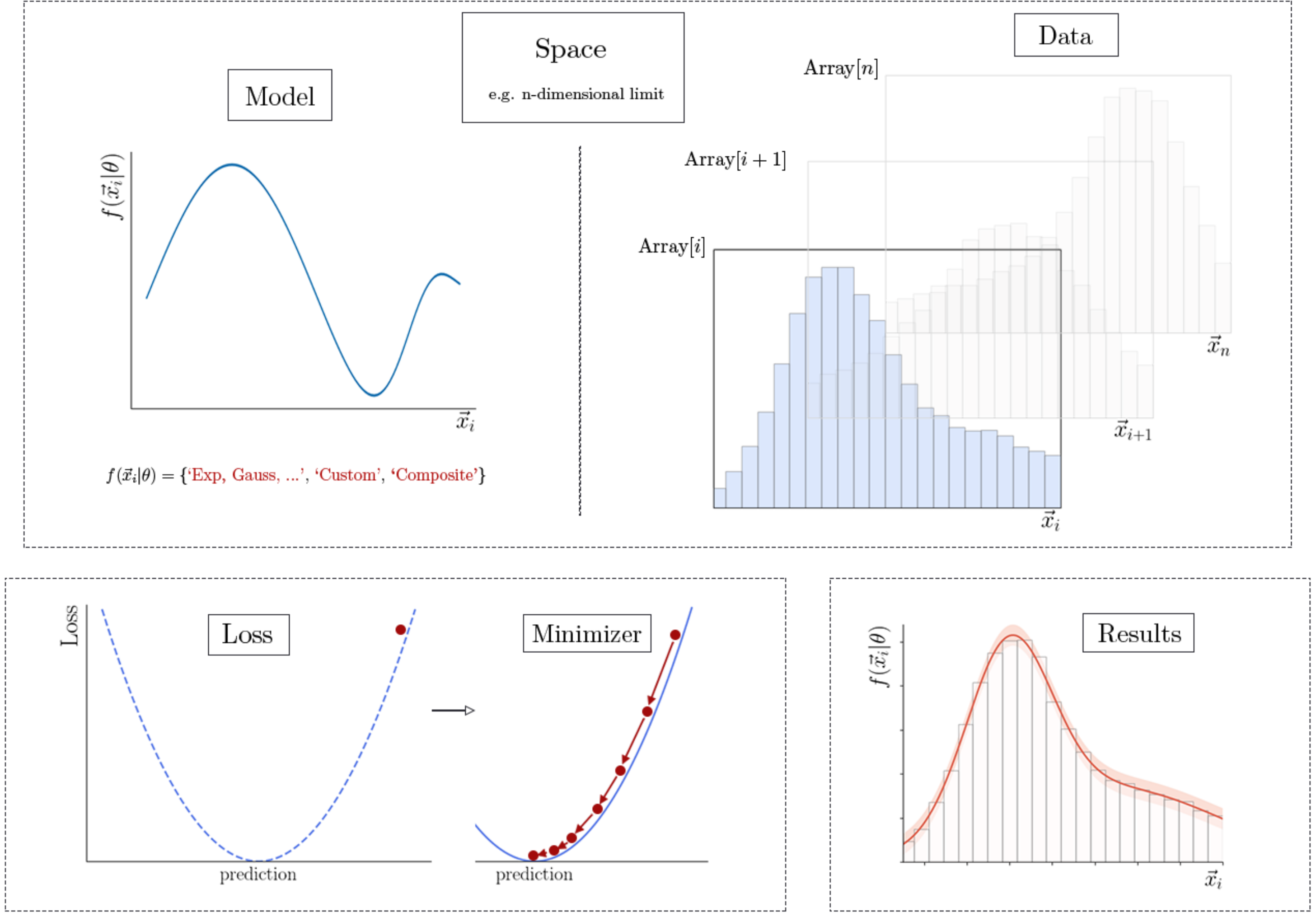}
    \caption{Schematic zfit workflow view of the individual structural blocks.\label{fig:sw-scheme}}
    \label{fig:fit_workflow}
\end{figure}

The structure of a generic model fitting workflow is illustrated in \fig{sw-scheme}.
Its basic goal is to maximize the agreement of a set of data points to a probability distribution described by a set of parameters.
The measurement of the metric of the goodness---or, more precisely, the disagreement---of the predicted \emph{model} with the experimental \emph{data} is referred to as \emph{loss} function.
One of the most used loss functions in HEP is based on the likelihood function, in particular the negative log-likelihood, which is  minimized in order to estimate the parameters of the model. 
Once the optimal values of the model parameters have been determined, several statistical techniques can be used to calculate their \emph{uncertainties}.

The structure defined by \emph{model}-\emph{data}-\emph{loss}-\emph{minimize} can also be used to describe a typical Deep Learning workflow---even more clearly by replacing \emph{model} by ``Neural Network'' and \emph{minimize} by ``Training''.
This similarity has profound implications in the design of the zfit implementation, as it allows to use an industry standard Deep Learning tool such as TensorFlow~\cite{tensorflow2015-whitepaper} (TF) as its computational backend by using the low-level API that consists of mathematical functions.
The immediate benefits for zfit in terms of performance arise from key features of TF such as the capacity to be run on heterogeneous architectures, mathematical optimizations such as automatic differentiation (required by the most used minimization algorithms in HEP), and operation caching. This puts the performance of zfit in the same order of magnitude as the existing standard as shown in~\ref{appendix:performance}.

zfit does not intend to be a large, monolithic package and has therefore a clear, restricted scope:
model fitting and sampling with a well defined and consistent API which is mostly independent of the implementation.
Tasks such as data processing, plotting and statistical analysis are left for other packages to tackle.

The abstraction of the fitting process in the steps shown in \fig{sw-scheme} is one of the main features of the architecture of zfit:
by developing each step as a separate ``brick'' of the library, it becomes possible to tightly integrate with the scientific computing Python ecosystem, and use the available tools to improve performance, reduce code duplication and add extra functionality.
Each of these bricks (\emph{model}, \emph{data}, \emph{loss}, \emph{minimizer} and \emph{fit result}) provides override hooks for all its core functionality, allowing to completely customize or replace any aspect of their behavior.
As hinted at above, these object deal internally with TF. Since using TF in the correct way adds another layer of difficulty, most of these additional complications are hidden from the user and the library offers a similar user experience as the one found in other model fitting libraries, independent of the actual backend.

Model fitting handles multiple dimensions. There are observables, the dependent variables, over which a model is usually normalized and parameters, which will be adjusted during the minimization of the loss. In contrast to other frameworks such as RooFit, zfit makes a strong distinction between observables and parameters; this removes any ambiguity in terms of normalization, allows to be closer to Deep Learning and the optimized way of how TF works.
The observables and corresponding limits in which the model fit is performed are handled by the \emph{Space} object.
Space objects can be single- or multi-dimensional, and, thanks to the use of observables, can be combined using the Python arithmetic operators to tackle arbitrarily complex uses cases.

Models are defined inside a given Space and implement a shape function that returns a value given some input data.
Custom models are easily implemented using their functional form and by default receive standard attributes such as numerical integration and sampling.
These generic functionalities can also be extended to analytical integration and specific sampling if provided by the user. 
Models can be combined to build more complex models (including increasing the number of dimensions) while keeping automatic integration, observable handling and event sampling. The normalization is done over a range specified in the \pyth{obs} or further customized by overriding the \pyth{_normalization} method manually.

Models are parametrized by \emph{Parameter} objects, which can be used in the implementation of the shape function like any other scalar.
\emph{Independent parameters}
are the only type of objects that are left free in the fit.
\emph{Composed parameters} can be built by combining parameters through mathematical operators and can be used like independent parameters.

The loading, ordering and processing of data in zfit is handled uniformly, no matter their source, by the \emph{Data} object.
While only the most important formats are currently implemented---ROOT trees, Numpy arrays and Pandas DataFrames, as well as TF tensors---the simple API of the Data object makes it easy to add additional data handling capabilities.

Data and Models are combined to build a \emph{Loss} function, which can either be pre-defined such as the negative log-likelihood or completely arbitrary. The computation of the value of the loss for a given configuration of the model is efficiently implemented in zfit thanks to TF.
It is also possible to add constraints to the loss in order to modify its behavior. Constraints are additional terms to the likelihood that often represent some external information on the parameter. While the most used constraints in HEP are directly implemented in zfit, such as Gaussian constraints,
arbitrary custom functions can be also used.

Minimization is performed through stateless Minimizer objects.
Since the loss calculation is usually the most intensive part of the minimization problem, in practice it is possible to wrap any existing Python minimizer to be used by zfit.
Several minimizers are included in zfit, such as the SciPy optimizers, the most-used minimizer in HEP, called Minuit~\cite{minuit}, or the TF implementation of the Adam optimizer~\cite{adam}.

The results of the minimization process are returned in a FitResult object, which stores all the information about the fit parameters, the minimization process, as well as the used instances of the Minimizer and the Loss. This includes the function minimum and the covariance matrix.

While the approximate uncertainties related to the fit parameters can be calculated directly inside zfit using the Hessian matrix or a simple profiling method, the calculation of precise parameter uncertainties and confidence intervals are left to specialized packages, such as those in Refs.~\cite{lauztat,hepstats},  which don't need any object other than the FitResult.

\section{Illustrative examples}
\label{sec:examples}

The architecture and features of zfit are better visualized with a typical HEP example:
a fit to the invariant mass of multiple particles including both signal and background events, extended to multiple dimensions,
and introducing a control channel.

\subsection{Basic usage}

Let's assume the domain of interest of the mass observable, in this case the $B^0$ meson mass, is in the range $[4.5, 6.0]\,$GeV.
In zfit this is expressed with a \texttt{Space} defining our domain:
\begin{center}
\begin{minipage}{\textwidth}
\begin{python}
obs = zfit.Space(obs="mass", limits=(4.5, 6.0))
\end{python}
\end{minipage}
\end{center}
The best interpretation of the observable at this stage is that it defines the name and range of the observable axis.

Data can be loaded for example from a numpy array, where the \texttt{obs} of the Space are matched by order with the columns of the array:
\begin{center}
\begin{minipage}{\textwidth}
\begin{python}
data = zfit.Data.from_numpy(array=numpy_array_data, obs=obs)
\end{python}
\end{minipage}
\end{center}
 with \pyth{numpy_array_data} being a numpy array holding our data. The signal component of the fit will be modelled using a Gaussian (normal), which has two free parameters---its mean and its width:
\begin{center}
\begin{minipage}{\textwidth}
\begin{python}
mu = zfit.Parameter("mu", 5.2, 4.5, 6.0)
sigma = zfit.Parameter("sigma", 20, 1, 40)
\end{python}
\end{minipage}
\end{center}
where the arguments for the parameters are: a unique name, an initial value, a lower and an upper bound. The bounds are not mandatory, the parameter will be floating in both cases. These can be used  to instantiate the Model as:
\begin{center}
\begin{minipage}{\textwidth}
\begin{python}
signal = zfit.pdf.Gauss(obs=obs, mu=mu, sigma=sigma)
\end{python}
\end{minipage}
\end{center}
 
While most commonly-used models are provided within the \texttt{pdf} module for convenience, the philosophy of zfit is to provide a clear API with powerful mechanisms to facilitate that the users and the community can implement their own.
Making use of another built-in model, the background that is parametrized as an exponential:
\begin{center}
\begin{minipage}{\textwidth}
\begin{python}
bkg = zfit.pdf.Exponential(-0.001, obs=obs)
\end{python}
\end{minipage}
\end{center}
 
In order to build the sum of these two models, an additional parameter \texttt{frac} is used to describe the fraction of each species:\footnote{Following a similar approach to the reference implementation of RooFit, the fraction of the last species corresponds to $1-\sum\text{frac}_i$.}
\begin{center}
\begin{minipage}{\textwidth}
\begin{python}
frac_sig = zfit.Parameter("signal fraction", 0.5, 0, 1)
model = zfit.pdf.SumPDF(pdfs=[signal, bkg], fracs=frac_sig)
\end{python}
\end{minipage}
\end{center}

With the model and the data built, a loss function can be constructed.
In this case, the goal is to perform an unbinned maximum likelihood fit, so the corresponding Loss object is initialized as
\begin{center}
\begin{minipage}{\textwidth}
\begin{python}
nll = zfit.loss.UnbinnedNLL(model=model, data=data)
\end{python}
\end{minipage}
\end{center}
 
Finally, the minimization process is executed, first instantiating a Minimizer object---in this case using the Minuit minimizer---and then running the minimization algorithm:
\begin{center}
\begin{minipage}{\textwidth}
\begin{python}
minimizer = zfit.minimize.MinuitMinimizer()
result = minimizer.minimize(nll)
\end{python}
\end{minipage}
\end{center}

The outcome of this minimization is stored in a \pyth{FitResult} object, 
which provides access to all details that occurred during the convergence process and to the best values estimated for the free parameters together with a nice printout of the parameters and uncertainties.
For example, one could obtain the fitted value of the signal mean as\footnote{Notice 
that 
\pyth{mu}, the parameter object, and not \pyth{"mu"}, the name of it, is used 
as the key.}
\begin{center}
\begin{minipage}{\textwidth}
\begin{python}
if result.converged:
    mu_value = result.params[mu]["value"]
\end{python}
\end{minipage}
\end{center}

The \pyth{params} object stores additional information such as estimated uncertainties if they have been calculated.

\subsection{Advanced usage}

The next step is to extend the model to take into account extra dimensions of the problem, in this case the angular distributions of the decay products of the $B^0$ meson.
To do so, the signal distribution is modelled with a custom model, which is created by inheriting from one of the base classes provided by zfit:
\begin{center}
\begin{minipage}{\textwidth}
\begin{python}
from zfit import z

class AngularSignalPDF(zfit.pdf.ZPDF):
    _PARAMS = ["FH", "AFB"]
    _N_OBS = 1
    
    def _unnormalized_pdf(self, x):
        # retrieve data
        costhetal = z.unstack_x(x)
        # retrieve parameters
        FH = self.params["FH"]
        AFB = self.params['AFB']
        # build the shape
        pdf = (3 / 4 * (1 - FH) * (1 - z.square(costhetal))
               + 0.5 * FH + AFB * costhetal)
        return pdf
\end{python}
\end{minipage}
\end{center}
 
Instead of \pyth{BasePDF}, we use here an even simpler class, \pyth{ZPDF}, which only requires to specify the naming of the parametrization via the \pyth{_PARAMS} attribute and allows, optionally, to define the dimensionality through \pyth{_N_OBS}. The latter also allow to create n-dimensional Models as \pyth{x} will be a list of length \pyth{_N_OBS} containing the data. Internally, the class handles the management of the free parameters, which can be accessed by their parameterization name through the \pyth{params} attribute, which holds the parameters. It also provides a simple way to access the input data with the \pyth{z.unstack_x}.
The module \pyth{z} provides some TF functions wrapped for better handling of dtypes as well as
additional methods and can be used mixed with pure TF functions.

\begin{center}
\begin{minipage}{\textwidth}
\begin{python}
# create the parameters
fh = zfit.Parameter("FH", 1.)
afb = zfit.Parameter("AFB", 0.5)

obs_costhetal = zfit.Space("thetal", limits=(-1, 1))

angular = AngularSignalPDF(obs=obs_costhetal, FH=fl, AFB=afb)
\end{python}
\end{minipage}
\end{center}

To create a two-dimensional model, one way is to multiply the models describing the different dimensions.
Given that their observables are different, the new model will automatically be defined in the combined space of the two:
\begin{center}
\begin{minipage}{\textwidth}
\begin{python}
combined_model = angular * model
\end{python}
\end{minipage}
\end{center}

This example can be extended to simultaneously fit a control dataset from which a more precise determination of the parameter \texttt{mu} can be obtained. 
To do so, a second Gaussian is created to model this control mode:
\begin{center}
\begin{minipage}{\textwidth}
\begin{python}
obs_control = zfit.Space("control", limits=(5.2, 5.6))
sigma_control = zfit.Parameter("sigma control channel", 2, 0, 5)
gauss_control = zfit.pdf.Gauss(obs=obs_control,
                               mu=mu  # common mu
                               sigma=sigma_control)
\end{python}
\end{minipage}
\end{center}
A simultaneous \texttt{Loss} can be built by either two independent \texttt{Loss} functions or by providing both models and datasets in a list as 
\begin{center}
\begin{minipage}{\textwidth}
\begin{python}
nll = zfit.loss.UnbinnedNLL([combined_model, gauss_control],
                            [data,           data_control])

\end{python}
\end{minipage}
\end{center}
where \texttt{data\_control} has been created analogously to \texttt{data}.
This simultaneous Loss can be minimized as shown before.

\section{Impact and conclusions\label{sec:impact}}

The zfit package is a crucial piece required to complete the transition of HEP analysis to a stack fully based on the scientific computing Python ecosystem.
The importance of this paradigm change should not be underestimated:
given the limited availability of resources in HEP, especially in areas such as software development, the possibility of integrating physics analysis within a well-developed and well-supported software ecosystem---allowing to focus mainly on the HEP-specific parts of the analysis software---will have long term repercussions in the quality and usability of HEP analysis software.

The package itself is not only an implementation of a versatile and powerful model fitting library, but also specifies a carefully designed and well discussed API of the most crucial components and the workflow. With the aim to be a standard in its kind, other packages such as high-level statistics and plotting tools, can be built against the standardized API instead of a concrete implementation, strongly reducing package dependencies. In addition, other packages can focus on the specific implementation of a certain task, such as building a specialized Loss, Model or a Minimizer, without the need to also implement the other parts of the workflow themselves.

% Another key feature of zfit is the ability to easily build custom models while being able to adjust the behavior of methods in an assisted or even fully user-controlled way. This core design feature allows the fast and safe implementation of arbitrary complex, custom models and therefore promotes
% the synergy between theory/experimental communities as illustrated in Refs.~\cite{Chrzaszcz:2018yza,Mauri:2018vbg}. 

Finally, fits in HEP analysis often consist of large datasets and complicated models, and require far more computing power than a single core. Scalability in terms of size and complexity of the data and models and the possibility to run on large computing infrastructures, including accelerators such as GPUs, is therefore a necessity. TensorFlow, the graph based computing backend in zfit, was built for the sole purpose of efficient large scale computing and takes care of the non-trivial task of optimal workload parallelization. This does not only yield state-of-the-art performance, but also enables the HEP community to keep up with the latest algorithm implementations and low-level computing instructions at virtually no cost.

In summary, a new high-level statistical modeling and fitting library purely designed within, and for, the Python ecosystem is presented. 
Future plans for zfit include full support to binned models, 
further extension on baseline HEP features as well as standardized, human-readable serialization of objects.

% To summarize, this paper has presented zfit, a high-level statistical modeling and fitting library purely designed within, and for, the Python ecosystem. 
% Its modular architecture and API allows users to not only formulate complex analyses within a simple formalism, but also easily to integrate developments from the scientific Python community and seamlessly use heterogeneous architectures.
% In light of the increasing and unprecedented statistics that will be collected in HEP experiments in the coming years, zfit will enable more rapid prototyping and will allow more efficient usage of the limited resources within academia.
% This will result in the completion of a paradigm change in data analysis that started with the eclosion of the scientific Python ecosystem.

\newpage

\section*{Acknowledgements}
We are grateful to Anton Poluektov, Chris Burr and Igor Babuschkin for demonstrating the potential of unbinned model fitting within the context of TensorFlow, which inspired this work. We also thank the Zurich LHCb Group, Matthieu Marinangeli, Josh Bendavid, Lukas Heinrich and the HSF community, especially Scikit-HEP project members, for useful discussions.  
A. Puig, R. Silva Coutinho, J.Eschle and N. Serra gratefully acknowledge the 
support by the Swiss National Science Foundation (SNF) under contracts 168169, 
174182 and 182622.

\newpage

\section*{Appendices}

\appendix

\section{Performance}
\label{appendix:performance}

A series of studies are performed to evaluate the execution time of zfit in comparison to the conventional RooFit library. 
The performance relies on two aspects, \textit{i.e.}
the complexity of the fit (given by the number 
of free parameters) and the data sample size. 
In this study, the sum of 9 Gaussian functions with a common mean and width parameter, but with different
central values, is used as baseline model.

Ensembles of Monte Carlo pseudoexperiments are generated and fitted with different setup 
configurations~\footnote{
The fitting performance is examined using a 12 core Intel i7 8850H with 2.60 GHz and 32Gb RAM, and a Nvidia P1000 with 4GB RAM.}.
\begin{enumerate}
	\item[i.] Nominal zfit implementation on CPU, using an 
	analytic gradient provided by TF. In this case, an initial run is done to 
	remove the graph compiler time. While not significant, this provides a more 
	realistic estimation.
	\item[ii.] Alternatively the 
	analytic gradient provided by TF is disabled, denoted by the addition 
	``nograd''. Instead, the Minuit minimizer 
	calculates a numerical approximation of the gradient internally.
	\item[iii.] The default CPU zfit settings is replaced by the GPU implementation. 
	\item[iv.] A standard RooFit implementation using the Python bindings with 
	PyROOT (version 6.16.00). The parallelization is done when invoking the \pyth{fitTo} method 
	and equals the number of cores available.
\end{enumerate}
For each setup, twenty toys are generated with sample sizes from 128 to 8 million events (except for the
GPU, where it goes only up to 4 million\footnote{The GPU used in these tests has a relative small memory. 
Performing larger-than-memory computations and multi-GPU is still work in 
progress.}) and the average fitting time is used as reference point. 
Note that for simplicity the Minuit algorithm is used in all cases. 
\begin{figure}[tbp]
\centering
\includegraphics[width=0.6\textwidth]{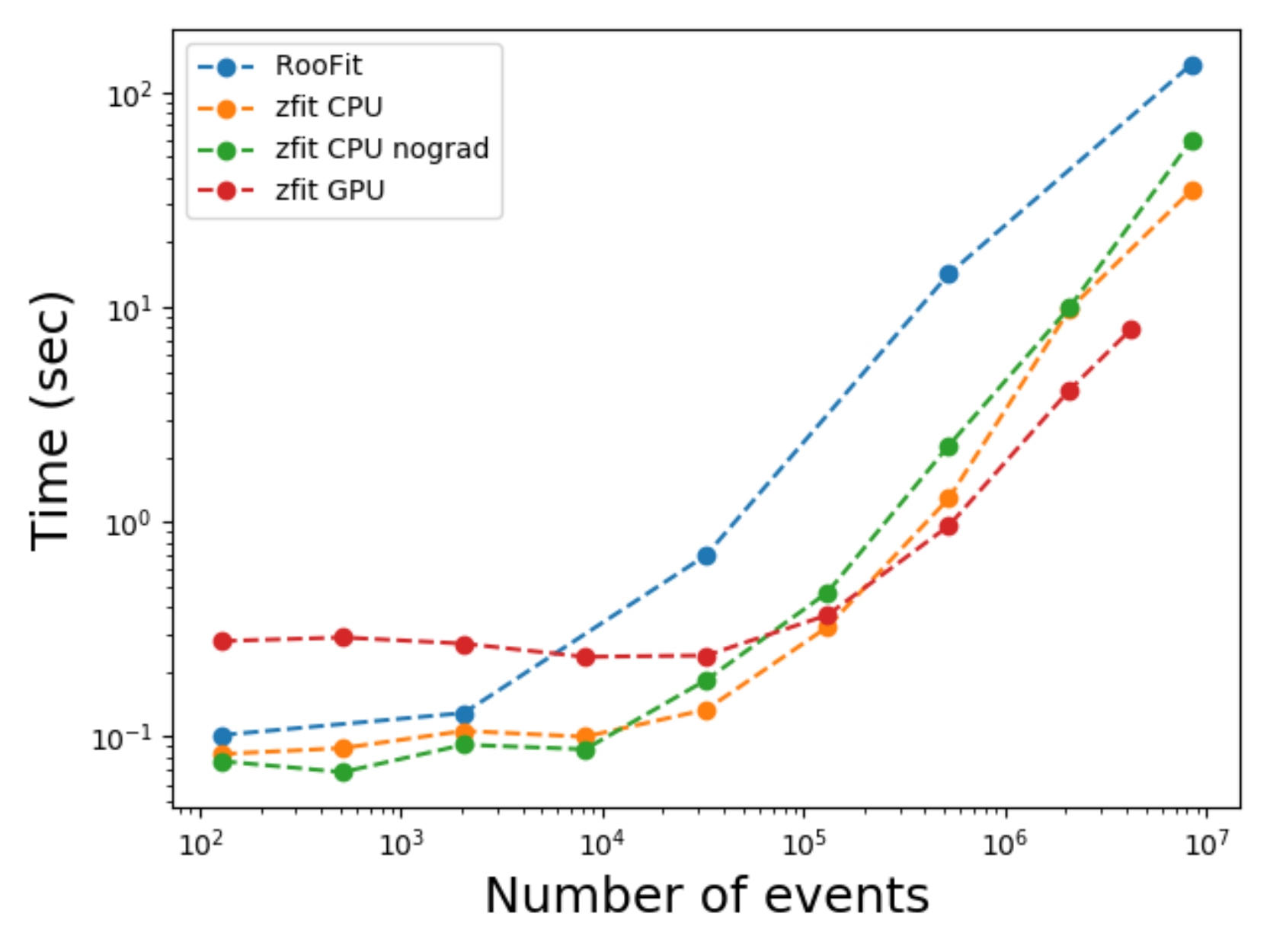}
\caption{Averaged time per fit for a sum of Gaussian  with shared mean and width as a function of the the number of generated events.}
\label{fig:gperf}
\end{figure}

Figure~\ref{fig:gperf} compares the performance of different fit scenarios and can be summarize as follows.
\begin{itemize}[-]
	\item As the number of events increase, the execution time of RooFit 
	monotonically increases.~\footnote{It is worth mentioning that the RooFit parallelization strategy is rather simple and with very few overheads; several improvements in this direction are ongoing. It however hints an important point of using TF in zfit, namely that this kind of work is factored out and a superior performance can come for free.}
	\item There is no advantage using parallelization for 
	very few calculations, since the overhead of splitting and collecting the 
	results is dominant. 
	With increasing number of events this gets negligible and thus  
	the execution time of zfit 
	increases way slower than that for RooFit. 
	This also comes from the fact that 
	more events mean a more stable loss shape, so 
	the minimizer used in zfit performs better.
	\item The GPU is highly efficient in computing 
	thousands of events in parallel. For only a few data points though, the overhead of moving data back and forth dominates, 
	resulting in a worse performance in this regime.
\end{itemize}

As a conclusion, the speed for a multicore CPU system is comparable between zfit and RooFit. 
For large number of events, zfit typically outperforms with a similar response for both GPU and a multicore system. 

\newpage

\section*{Required Metadata}

\section*{Current code version}
\label{sec:required-code}

\begin{table}[!h]
\begin{tabular}{|l|p{5.5cm}|p{6.5cm}|}
\hline
\textbf{Nr.} & \textbf{Code metadata description} &  \\
\hline
C1 & Current code version & 0.3.6 \\
\hline
C2 & Permanent link to code/repository used for this code version & $https://github.com/zfit/zfit$ \\
\hline
C3 & Legal Code License   & BSD-3 \\
\hline
C4 & Code versioning system used & Git \\
\hline
C5 & Software code languages, tools, and services used & Python, TensorFlow \\
\hline
C6 & Compilation requirements, operating environments \& dependencies & Linux and MacOS supported (Windows functionality unknown); \newline Python 3.6+; \newline
dependencies (also specified in $requirements.txt$): \newline $tensorflow>=1.14.0,<2,$ \newline
$tensorflow\_probability>=0.6.0,<0.8,$ \newline
$scipy>=1.2$ \newline
$uproot,$
$pandas,$
$numpy,$
$iminuit,$\newline
$typing,$
$colorlog,$
$texttable,$
$ordered$-$set$
\\
\hline
C7 & If available Link to developer documentation/manual & $https://zfit.readthedocs.io/en/0.3.6/$ \\
\hline
C8 & Support email for questions & zfit@physik.uzh.ch \\
\hline
\end{tabular}
\caption{Code metadata.}
\label{} 
\end{table}

\newpage

\section*{Current executable software version}
\label{sec:current_exec}

\begin{table}[!h]
\begin{tabular}{|l|p{4.5cm}|p{8.0cm}|}
\hline
\textbf{Nr.} & \textbf{Executable software metadata description} &  \\
\hline
S1 & Current software version & 0.3.6 \\
\hline
S2 & Permanent link to executables of this version  & $https://github.com/zfit/zfit/releases/tag/0.3.6$ \\
\hline
S3 & Legal Software License & BSD-3 \\
\hline
S4 & Computing platforms/Operating Systems & Unix-like (Windows untested) \\
\hline
S5 & Installation requirements \& dependencies & 
dependencies (also specified in $requirements.txt$): \newline $tensorflow>=1.14.0,<2,$ \newline
$tensorflow\_probability>=0.6.0,<0.8,$ \newline
$scipy>=1.2$ \newline
$uproot,$
$pandas,$
$numpy,$
$iminuit,$\newline
$typing,$
$colorlog,$
$texttable,$
$ordered-set$
\\
\hline
S6 & If available, link to user manual - if formally published include a reference to the publication in the reference list & $https://zfit.readthedocs.io/en/0.3.6/$ \\
\hline
S7 & Support email for questions & zfit@physik.uzh.ch\\
\hline
\end{tabular}
\caption{Software metadata}
\label{} 
\end{table}

\newpage

\bibliographystyle{elsarticle-num} 
\bibliography{main.bib}

\end{document}